\documentclass[aps,pra,superscriptaddress,showpacs,twocolumn]{revtex4-2}
\usepackage{bm}
\usepackage{amsmath}
\usepackage{amssymb}
\usepackage{slashed}
\usepackage{float}
\allowdisplaybreaks
\usepackage{subcaption}
\usepackage{dcolumn}
\usepackage{graphicx}
\newcolumntype{.}{D{x}{}{-1}}

\newcolumntype{w}[1]{D{.}{.}{#1}}
\newcolumntype{L}{>{$}l<{$}}
\newcommand{\Za}{Z\alpha}
\usepackage{nicefrac}

\begin{document}

\title{Accurate determination of $^{6,7}$Li nuclear magnetic moments}

\author{Krzysztof Pachucki}
\affiliation{Faculty of Physics, University of Warsaw,
             Pasteura 5, 02-093 Warsaw, Poland}

\author{Vojt\v{e}ch Patk\'o\v{s}}
\affiliation{Faculty of Mathematics and Physics, Charles University,  Ke Karlovu 3, 121 16 Prague
2, Czech Republic}

\author{Vladimir A. Yerokhin}
\affiliation{Max–Planck–Institut f\"ur Kernphysik, Saupfercheckweg 1, 69117 Heidelberg, Germany}

\begin{abstract}
We report an accurate determination of the nuclear magnetic dipole moments of $^{6,7}$Li
from the measured ratio of the nuclear and the electron $g$-factors
in atomic Li. The obtained results significantly improve upon the
literature values and stress the importance of reliable theoretical
calculations of the nuclear shielding corrections.
\end{abstract}

\maketitle

Standard tabulations of the nuclear moments \cite{raghavan:89,stone:05} often rely on data derived
from the nuclear magnetic resonance measurements. In order to extract the
magnetic moment of bare nuclei from these experiments, one has to correct for the so-called
chemical shifts originating from the surrounding electrons and neighbouring atoms.
Such shifts are difficult to calculate reliably, which often leads to significant systematic
uncertainties in the literature data on nuclear magnetic moments. As an example, the
so-called ``bismuth hyperfine puzzle'' \cite{ullmann:17,karr:17} was recently resolved
\cite{skripnikov:18}
and traced back to the previously underestimated uncertainty of the
correction due to the chemical surrounding in the determination of
the nuclear magnetic moment of bismuth.

Much more accurate determinations of nuclear magnetic moments can be performed from measurements
of the combined Zeeman and hyperfine structure of atomic levels.
The nuclear and electron $g$ factors in atomic systems can be nowadays calculated
within QED
to a very high accuracy, with reliable estimations of uncertainties due to
uncalculated higher-order effects. High-precision
determinations of nuclear magnetic moments were recently reported for
deuterium and tritium \cite{puchalski:22}, $^3$He \cite{wehrli:21}, and $^9$Be
\cite{pachucki:10}.
In this work we perform an accurate determination of the
nuclear magnetic dipole moments of $^{6,7}$Li from measured
ratios of the nuclear and the electron $g$-factors \cite{beckmann:74}.
The nuclear moment of $^7$Li is of particular importance since it is used
as a reference in determinations of magnetic moments of unstable $^{8,9,11}$Li isotopes
\cite{winnacker:78,borremans:05,neugart:08}.
Improved values of the nuclear magnetic moments of Li isotopes are
also needed in view of continuing efforts
 \cite{puchalski:13, sun:23, pachucki:23} to determine values of
the so-called effective Zemach radii characterizing the
magnetization distribution over the nuclear volume
from hyperfine-structure measurements.

Magnetic $g$-factors of an atomic system are defined through the effective
interaction with the magnetic field
\begin{equation}
H = -\frac{e}{2\,m_e}\,g_J\,\vec J\cdot\vec B
-\frac{e}{2\,m_p}\,g_I\,\vec I\cdot\vec B+A\,\vec J\cdot\vec I\,,
\end{equation}
where $\vec{B}$ is the external magnetic field, $\vec I$ and $\vec J$ are the
total angular momentum of the nucleus and electrons, respectively, $g_I$ and $g_J$
are the $g$ factors of the nucleus and electrons, respectively, $m_p$ and $m_e$
are the mass of the proton and the electron, respectively, $e$ is the elementary change, and
$A$ is hyperfine constant. We note that our definition of the nuclear $g$-factor
contains $m_p$, whereas a different definition $g'_I = (m_e/m_p)\,g_I$ is
also used in the literature.

The nuclear $g$-factor $g_I$ of an atom is connected to the free-nucleus $g$-factor
$g_N$ by the shielding constant $\sigma$,
\begin{equation}
g_I = g_N\,(1-\sigma)\,,
\end{equation}
which can be accurately calculated by the atomic theory. For a light atom,
the shielding constant $\sigma$ is effectively described by a double
expansion in powers of the fine-structure constant $\alpha$ and the
electron-to-nucleus mass ratio $m_e/m_N$,
\begin{equation}
\sigma =
\alpha^2\,\sigma^{(2)} + \alpha^4\,\sigma^{(4)}
+  \alpha^2\,\frac{m_{ e}}{m_{ N}}\sigma^{(2,1)}+\ldots \,. \label{07}
\end{equation}

The first term of this expansion, $\sigma^{(2)}$, is obtained from
the Ramsey nonrelativistic theory
of the magnetic shielding \cite{ramsey:50:dia}.
For atomic systems it has a very simple form
\begin{equation}
\sigma^{(2)} = \frac{1}{3}\,\sum_a\biggl\langle\frac{1}{r_a}\biggr\rangle\,,
\end{equation}
where the summation over $a$ runs over all electrons. This matrix
element was calculated with a high accuracy
by Yan \cite{yan:95:prl,yan:95:li}
with the Hylleraas basis set. The result
for the leading-order shielding contribution
(in the infinite nuclear mass limit) for Li is
\begin{equation}
\alpha^2\,\sigma^{(2)} = 101.499 \times  10^{-6}\,.\label{10}
\end{equation}

The relativistic shielding correction, $\sigma^{(4)}$, was recently calculated
for helium \cite{rudzinski:09,wehrli:21}. For Li, there have been
no calculations of this contribution so far.
Here, we estimate this correction on the basis of known
hydrogenic result.
For hydrogenic ions, the relativistic shielding constant can be
derived analytically in a closed form \cite{pyper:99:b,ivanov:09}.
After expanding in $\Za$, the hydrogenic (H) result reads
\begin{equation}
\sigma_\mathrm{H} = \frac{\alpha\,(\Za)}{3\,n^2}\biggl[1+
\frac{132\,n-35}{36\,n^2}\,(\Za)^2 \biggr] + O(\alpha^6),\label{11}
\end{equation}
where $n$ is the principal quantum number. A straightforward application
of this formula to the ground state of Li leads to the following result in the
leading order in $\alpha$:
\begin{equation}
\alpha^2\,\sigma^{(2)} = \frac{\alpha^2}{3}\biggl[2\,Z+\frac{Z-2}{2^2}\biggr] = 111 \times 10^{-6}\,,
\end{equation}
which deviates only by 10\% from the exact result of Eq.~(\ref{10}).
Such an agrement is not surprising because mostly core
electrons contribute to $\sigma$,
and for Li they do not differ significantly from the hydrogenic case.
Next, we apply Eq.~(\ref{11}) to estimate the relativistic shielding correction
for the helium atom. The second term of the $\alpha$ expansion of Eq.~(\ref{11})
yields
\begin{equation}
\alpha^4\,\sigma^{(4)}_\mathrm{He} = 2\,\alpha^4\,Z^3\,\frac{97}{108}\,,
\end{equation}
With $Z=2$, it gives $0.040\,750$,
which is by 30\% smaller than the exact result
$\alpha^4\,\sigma^{(4)}_\mathrm{He} = 0.052\,663\,1$
\cite{wehrli:21, rudzinski:09}.
Analogously, we estimate the relativistic shielding correction for Li as
\begin{equation}
\alpha^4\,\sigma^{(4)} = \frac{\alpha^4}{3}\,\biggl[2\,Z^3\,\frac{97}{36} +
(Z-2)^3\,\frac{229}{576}\biggr]\,.
\end{equation}
and ascribe a 30\% accuracy to this approximation. We thus obtain
\begin{equation}
\alpha^4\,\sigma^{(4)} = 0.138\,(41)\times 10^{-6}.
\end{equation}

\begin{table}[t]
\caption{The magnetic moments of light nuclei.}
\begin{ruledtabular}
\begin{tabular}{lw{0.10}c}
\multicolumn{1}{l}{Element} &
        \multicolumn{1}{c}{$\mu/\mu_N$}
        &
        \multicolumn{1}{c}{Ref.}
\\
\hline\\[-5pt]
$^2$H & 0.857\,438\,233\,8\,(26)   & \cite{puchalski:22}\\
$^3$H & 2.978\,962\,465\,0\,(59)   & \cite{puchalski:22}\\
$^3$He & -2.127\,625\,350\,0\,(17)   & \cite{schneider:22,wehrli:21}\\
$^6$Li & 0.822\,044\,63\,(37)   & this work\\
       & 0.822\,045\,7\,(50) & \cite{makulski:18}\\
       & 0.822\,043\,(3) & \cite{stone:19} \\
       & 0.822\,047\,3\,(6)     & \cite{beckmann:74}, as quoted in \cite{stone:05} \\
       & 0.822\,567\,(3)        & \cite{lutz:68}, as quoted in \cite{stone:05} \\[1ex]
$^7$Li & 3.256\,416\,19\,(57)   & this work\\
       & 3.256\,418\,(20)               & \cite{makulski:18}\\
       & 3.256\,407\,(12) & \cite{stone:19} \\
       & 3.256\,427\,(2)  & \cite{beckmann:74}, as quoted in \cite{stone:05} \\
       & 3.256\,462\,5\,(4)   & \cite{lutz:68}, as quoted in \cite{stone:05} \\
$^9$Be & -1.177\,431\,59\,(3)  & \cite{pachucki:10}\\
\end{tabular}
\end{ruledtabular}
\end{table}

The last potentially important correction to $\sigma$ is the one due to
the finite nuclear mass. It is given by the formula \cite{pachucki:08:recoil}
\begin{eqnarray}
\sigma^{(2,1)} &=& \frac{1}{3}\,\biggl\langle\sum_a\frac{1}{r_{a}}\,
\frac{1}{(E-H)'}\,p_N^2\biggr\rangle
+\frac{1}{3}\,\frac{(1-\tilde g_N)}{Z\,\tilde g_N}\,\langle p_N^2\rangle
\nonumber \\&&
\hspace*{-8ex}+\frac{1}{3}\,\biggl\langle
\bigl(\vec r_1\times\vec p_2 +\vec r_2\times\vec p_1\bigr)\,
\frac{1}{(E-H)}\,
\sum_{a}\,\frac{\vec r_{a}}{r_a^3}\times \vec p_a\biggr\rangle,\label{15}
 \end{eqnarray}
where
$\vec p_N = -\sum_a \vec p_a,$ and
\begin{equation}
\tilde g_N = \frac{m_{ N}}{Z\,m_{ p}}\,
\frac{\mu}{\mu_{  N}}\,\frac{1}{I}\,.
\end{equation}
Since $\sigma^{(2,1)}$ is small, it does not need to be calculated accurately.
We therefore neglect the so-called mass-polarization part $\sum_{a>b}\vec p_a\cdot\vec
p_b$ in the first and the second term in Eq.~(\ref{15}) and neglect the third term
completely. After these simplifications, we
obtain the approximate form of the finite nuclear-mass correction as
\begin{equation}
\sigma^{(2,1)} \approx -\frac{1}{3}\,\biggl\langle\sum_a\frac{1}{r_{a}}\biggr\rangle
-\frac{2}{3}\,\frac{(1-\tilde g_N)}{Z\,\tilde g_N}\,E,
\end{equation}
where $E = -7.478\,060\,323$
is the nonrelativistic energy of Li in atomic units \cite{puchalski:08:li}.
The numerical results for the recoil contribution are
small but not negligible,
\begin{equation}
\alpha^2\,\frac{m_{ e}}{m_{ N}}\sigma^{(2,1)} \approx \left\{
\begin{array}{ll}
 -0.012 \times 10^{-6}\,, & \; \;\mathrm{for}\;^6\mathrm{Li},\\
 -0.013 \times 10^{-6}\,, & \; \;\mathrm{for}\;^7\mathrm{Li}.
 \end{array}\right.
\end{equation}

All other corrections to the shielding constant in Li are much smaller
and can be neglected on the level of our present interest. Finally,
our total values for the shielding constant are
\begin{equation}
\sigma = \left\{
\begin{array}{ll}
  101.62\,(4)\times 10^{-6}\,,   & \; \;\mathrm{for}\;^6\mathrm{Li}\,\\
  101.62\,(4)\times 10^{-6}\,,   & \; \;\mathrm{for}\;^7\mathrm{Li}\,.
 \end{array}\right.
\end{equation}
The uncertainty of these values comes from the estimate of the relativistic correction;
this is the factor limiting the accuracy of our determination of nuclear moments.

The electron $g$ factor in Li was calculated by Yan in Refs.~\cite{yan:01:prl,yan:02:jpb}.
Recently, Shabaev et al.~\cite{shabaev:17:prl} corrected the nuclear recoil
part of Yan's calculation. Using Yan's values for the $\alpha^2$ and $\alpha^3$
corrections and the nuclear recoil values from Ref.~\cite{shabaev:17:prl},
we obtain
\begin{align}
\frac{g_J}{g_e} - 1  = \left\{
\begin{array}{ll}
 -9.124\,99\,(17) \times 10^{-6}\,, & \; \;\mathrm{for}\;^6\mathrm{Li},\\
 -9.125\,19\,(15) \times 10^{-6}\,, & \; \;\mathrm{for}\;^7\mathrm{Li},\\
 \end{array}\right.
\end{align}
where $g_e$ is the free-electron $g$ factor. The
uncertainties of the above values
come from the estimated errors of numerical computation in
Ref.~\cite{shabaev:17:prl}.

We are now ready to determine the nuclear magnetic moments of $^{6,7}$Li. They
are obtained from the measured ratio of the nuclear and electron $g$ factors
\cite{beckmann:74},
\begin{align}
g'_I/g_J = \left\{
\begin{array}{ll}
-2.235\,697\,8(10)\times 10^{-4}\,, \; \;\mathrm{for}\;^6\mathrm{Li}, \\
-5.904\,271\,9(10)\times 10^{-4}\,, \; \; \mathrm{for}\;^7\mathrm{Li},\\
\end{array}\right. \label{01}
\end{align}
by means of the formula
\begin{align}
\frac{\mu}{\mu_N} =\ g_N\,I
    = \frac{g'_I}{g_J}\,\frac{g_J}{g_e}\,
        \frac{g_e}{(1-\sigma)}\,\frac{m_p}{m_e}\,I\,.
\end{align}
By using the values for the shielding constant $\sigma$ and the electron $g$-factor $g_J$
as summarized above and the
free electron g-factor and physical constants from \cite{tiesinga:21:codata18,wang:12},
we obtain the results for the nuclear magnetic moments of $^{6,7}$Li
as presented in Table I.

The comparison with the literature data
given in Table I shows significant deviations of the present
nuclear magnetic moments from those tabulated in Ref.~\cite{stone:05}.
The reason is that the previous tabulations typically ignored
uncertainties associated with theoretical calculations of diamagnetic corrections.
This problem has been recognized and addressed in the updated tabulation \cite{stone:19}.
In particular, for $^{6,7}$Li, the nuclear moments from Ref.~\cite{stone:19}
have much larger uncertainties
than previously and are in agreement within the present values.

The improved values of the nuclear magnetic moments
influence the determinations of the Zemach radii $r_Z$
reported in Refs.~\cite{puchalski:13,sun:23}. Specifically,
the shift of our values of nuclear magnetic moments as compared to those of
Ref.~\cite{stone:05} leads to a change of $r_Z$ by 0.02~fm for $^6$Li and
0.03~fm for $^7$Li, which is comparable with the claimed uncertainties
of the $r_Z$ values in Refs.~\cite{puchalski:13,sun:23}.
The most recent values of the effective Zemach radii are
$\tilde r_z(^6\mathrm{Li}) = 2.39(2)$~fm,
and $\tilde r_z(^7\mathrm{Li}) = 3.33(3)$~fm \cite{pachucki:23}.

In summary, we have determined the nuclear magnetic dipole moments of $^{6,7}$Li with an improved
precision as compared to the literature values.
Our work indicates the importance of reliable estimations of uncertainties
in the calculation of corrections to the nonrelativistic nuclear magnetic shielding.

%\bibliographystyle{c:/-a-/papers/bibtex/phaip30}
%\bibliography{c:/-a-/papers/bibtex/hfst}

\end{document}